\documentclass[
 superscriptaddress,
 amsmath,amssymb,
 aps,
]{revtex4-2}

\usepackage{graphicx}
\usepackage{dcolumn}
\usepackage{bm}
\usepackage{hyperref}

\usepackage{xcolor}
\usepackage[section]{placeins}
\begin{document}

\preprint{}

\title{Biswas-Chatterjee-Sen kinetic exchange opinion model for two connected groups}

\author{Krzysztof Suchecki}
\affiliation{Faculty of Physics, Warsaw University of Technology, Koszykowa 75, 00-662 Warsaw, Poland}

\author{Kathakali Biswas}
\affiliation{Department of Physics, University of Calcutta, 92 Acharya Prafulla Chandra Road, Kolkata 700009, India}
\affiliation{Department of Physics, Victoria Institution (College), 78B Acharya Prafulla Chandra Road, Kolkata 700009, India}

\author{Janusz A. Ho\l yst}
\affiliation{Faculty of Physics, Warsaw University of Technology, Koszykowa 75, 00-662 Warsaw, Poland}

\author{Parongama Sen}
\affiliation{Department of Physics, University of Calcutta, 92 Acharya Prafulla Chandra Road, Kolkata 700009, India}

\date{\today}

\begin{abstract}
We consider a kinetic model of opinion dynamics known as the Biswas-Chatterjee-Sen model with a modular interaction structure.
The system consists of two groups of agents that feature more frequent interactions within each group and rarer interactions between agents of different groups.
We use the mean-field analytical approximation to determine that aside from previously known ordered and disordered states, a new antisymmetric ordered state is stable, where each group has an opposite dominant opinion.
The limits of system interaction strength and noise for the stability of such a state are determined, with a discontinuous transition from an antisymmetric to a symmetric state happening if thresholds are exceeded.
The results of numerical agent-based simulations confirm our analytical predictions and show that the critical values of noise and interaction strength are predicted with good accuracy.
\end{abstract}

\maketitle

\section{Introduction}
\label{sec-introduction}
Applications of statistical physics to understand social phenomena have attracted a lot of interest in recent times \cite{Castellano2009}.
The problem of opinion formation in a society with regard to public issues (e.g., Brexit) is possibly the most addressed one \cite{Castellano2009, SenChakrabarti2013,Galambook2016}.
While proposed models are far too simple to reproduce complex human behavior in detail, many models represent certain aspects of real opinion formation processes, showing a relation between assumed microscopic human behavior and emergent phenomena observed at the scale of the entire social group.
Of special interest is the issue of whether a model can reach a stable \emph{consensus}, where all individuals share the same opinion, or at least a similar stable state where the majority shares it.
The question is, under what assumptions of microscopic behavior, initial and environmental conditions the consensus or stable majority is reached, and under what conditions it is not.
The consensus and  similar states, where the majority shares the same opinion, are often considered ordered states, while situations where opinions of individuals appear or are, in fact, random are considered disordered states.
The appearance or disappearance of consensus with changing model parameters can be, therefore, understood as an order-disorder transition, similar, for example, to order-disorder transition in interacting spin systems such as the Ising model \cite{dorogovtsev2002ising,macy2024ising}.

Inspired by the kinetic theory of gases, a class of models for socio-economic systems has been proposed \cite{toscani2006, lallouache2010} called kinetic exchange models.
In these models of opinion formation, agents interact with other agents in discrete events similar to collisions between gas particles, and get their opinions updated according to a given rule.
The opinion values can be defined to be discrete or continuous. 
When using discrete opinions, these models are known \cite{biswas2012,mukherjee2016} to display a behavior similar to majority models or spin models \cite{Castellano2009}.
It is also well known that depending on how interactions between individuals are structured, many different models may display different behaviors or properties.
Examples include: a) the critical temperature of the Ising model in finite scale-free networks which depends on the system size \cite{aleksiejuk2002, bianconi2002} as opposed to being size-independent for lattices, random graphs or well-mixed spin systems described by a mean-field; b) absence or presence of epidemic threshold in models of epidemic spreading depending on whether the interaction structure is a scale-free network or not and whether it features local clustering of interactions \cite{pastorsatorras2001,eguiluz2002}; and c) existence of cascading failures in multi-layered networks \cite{buldyrev2010} that would not propagate if networks were separate.

Real societies often feature a community structure, where people belonging to the same group interact strongly or frequently, while people that are from different groups interact weakly or less often.
This corresponds to a modular network structure with internally densely linked modules that are sparsely linked in between.
Both the Ising model \cite{suchecki2006, suchecki2009, bolfe2018} and the majority model \cite{lambiotte2007} are known to feature an additional ordered state other than the consensus if a community structure is present.
In this additional state, the interior of each group is ordered, with agents or spins sharing mostly the same opinion or state, and the mean opinion or state of each group is different.
In the case of two groups and binary state models, as is the focus of this manuscript, we will call this type of state an antisymmetric ordered state.
Similar system states can also be exhibited by continuous opinion models in modular environments \cite{gajewski2022transitions}.
Given some similarity of the behavior between such models and discrete kinetic exchange models, our research question is whether the kinetic exchange models of opinion formation behave in similar way if we impose a modular interaction structure or not.

We have considered the BChS model \cite{BChSreview2023} belonging to the general class of kinetic exchange models where the total population is divided into two groups, and the probability of interaction between two agents depends on whether they belong to the same group or not.
Like in previous studies of this model, with probability $1-p$ the interactions are attractive, bringing opinions of interacting agents closer, or with the probability $p$ they are repulsive, driving the opinions of the interacting agents apart.
In this work, we add a new parameter $\alpha$, which determines the probability for agents to interact with agents from another group instead of their own.
We are interested to see whether an antisymmetric ordered state exists, is stable, and how correlated are the average opinions of communities. 
Our results show that there are no stable ordered states for $p>\tfrac{1}{4}$, as in the original BChS model, for any value of the mixing probability $\alpha$.
For $p<\tfrac{1}{4}$, we find that the ordered phase can be either symmetric (both groups have average opinions of the same sign) or antisymmetric depending on the values of $p$ and $\alpha$.
This means that two types of transitions can occur in modular systems -- a typical continuous order-disorder phase transition as well as a discontinuous transition between two different ordered states, similar to what happens in modular spin systems \cite{suchecki2009}.

\section{Model}
\label{sec-model}
The BChS model \cite{biswas2012} is investigated with discrete opinions of the agents.
The opinion of the $i$th agent is written as  $o_i \in \{ -1,0,1\}$.
The opinions of the agents evolve in time.
Opinion $o_i$ is updated due to her interaction with a randomly chosen agent $j$ as
\begin{equation}
	o_i(t+1)=o_i(t)+\mu_{ij}(t) o_j(t) \label{eq-oit}.
\end{equation}
The interaction $\mu_{ij}$ is assumed to be $1$ with probability $1-p$ and $-1$ with probability $p$, where $p$ is a constant model parameter expressing noise inherent to the dynamics, from a deterministic system at $p=0$ to completely random at $p=\tfrac{1}{2}$.
The probability of repulsive interactions $p$ can be also interpreted as a probability of contrarian or non-conformist behavior and fulfills a similar role to temperature \cite{dorogovtsev2002ising}, with $p=1/2$ corresponding to $T=+\infty$.
The opinions are bounded, so any result of $o_i(t+1)$ outside $\{-1,0,1\}$, such as due to interaction between two agents with $+1$ opinions and $\mu=1$, is truncated to the nearest allowed value.\\
Unlike in previous studies that considered well-mixed populations \cite{biswas2012} or considered an explicit network of interactions in the form of a graph \cite{alves2021,raquel2022}, we consider modular interactions, with agents belonging to two fixed groups $A$ and $B$ of the same size ($N_A=N_B$).
With probability $1-\alpha$ the interaction will occur with another agent from within the same group, and with probability $\alpha$ the interaction will be with an agent from the other group.
This corresponds to an annealed modular network with two modules, where fraction $\alpha$ of each vertex edges lead to the other module, with the rest leading to the same module.
For $\alpha=0$ the two groups are separate and do not interact, while for $\alpha=\tfrac{1}{2}$ the network does not possess a modular structure.
While $\alpha>\tfrac{1}{2}$ has a clear interpretation of a network with an increasingly bipartite structure, with $\alpha=1$ corresponding to a true bipartite network, we do not consider such situations and keep $\alpha \in (0,1/2)$.
Note that the groups represent interaction structure and are not determined by agent opinions, although ordered states may show a significant correlation between the opinions of agents and the groups they belong to.

\section{Analytical approach}
\label{sec-analytics}
To describe the system analytically, we follow a method similar to one established in \cite{biswas2012}, by considering fractions of agents with a specific opinion.
The difference is, instead of having just one group and fractions $f_-$, $f_0$ and $f_+$ representing densities of negative, neutral and positive opinions, we have two groups and consider fractions of each opinion in each of the groups $f_{A-}$, $f_{A0}$, $f_{A+}$, $f_{B-}$, $f_{B0}$, $f_{B+}$.
To determine the expected changes of these fractions, we consider the probabilities of interactions of two agents, considering both the parameters $p$ and $\alpha$. 
The agent with whom an interaction takes place is selected randomly from one's own group with probability $1-\alpha$ or from the other group with probability $\alpha$ (as defined in Sec \ref{sec-model}) and the interaction can be repulsive or attractive with probability $p$ and $(1-p)$ respectively in both cases.
We treat the probability of interacting with agent's own or the other group and the probability of interacting attractively or repulsively as independent.
This means that the probability of an agent interacting with an agent of a specific group in a specific way is a product of individual probabilities.
For example, the probability of an agent from group A, with opinion $o_i=+1$ interacting with an agent from group B with opinion $o_j=-1$ from the same group attractively ($\mu=+1$) is simply $f_{A+} \cdot f_{B-} \cdot \alpha \cdot (1-p)$.
Because the total number of agents in each group $A$ and $B$ is constant, the actual number of independent system variables is $4$ instead of $6$.
Following \cite{biswas2012} we elected to describe the state by \emph{mean opinions} $O_A=f_{A+}-f_{A-}$, $O_B=f_{B+}-f_{B-}$ and fractions of agents with neutral opinions $f_{A0}$, $f_{B0}$.
The resulting equations for changes of these variables can be written as
\begin{align}
    \frac{dO_A}{dt} &=O_A \left[ (1-\alpha) \left( (1-p) f_{A0} - p\right) -\alpha \frac{1-f_{B0}}{2}\right] + O_B \alpha (1-2p)\frac{1+f_{A0}}{2} \label{eq-dtoa}\\
    \frac{df_{A0}}{dt} &= (1-\alpha) \left[ 2(1-2p) \left( \frac{1-f_{A0}+O_A}{2}\right) \left( \frac{1-f_{A0}-O_A}{2}\right) + p (1-f_{A0})^2 -f_{A0}(1-f_{A0}) \right]+ \nonumber\\
    &+ \alpha \left[ (1-2p) \left( \frac{1-f_{A0}+O_A}{2}\right) \left( \frac{1-f_{B0}-O_B}{2}\right) + (1-2p) \left( \frac{1-f_{A0}-O_A}{2}\right) \left( \frac{1-f_{B0}+O_B}{2}\right) \right]+ \nonumber\\
    &+ \alpha \left[p(1-f_{A0})(1-f_{B0})-f_{A0}(1-f_{B0}) \right] \label{eq-dtfaz}\\
    \frac{dO_B}{dt} &=O_B \left[ (1-\alpha) \left( (1-p) f_{B0} - p\right) -\alpha \frac{1-f_{A0}}{2}\right] + O_A \alpha (1-2p)\frac{1+f_{B0}}{2} \label{eq-dtob}\\
    \frac{df_{B0}}{dt} &= (1-\alpha) \left[ 2(1-2p) \left( \frac{1-f_{B0}+O_B}{2}\right) \left( \frac{1-f_{B0}-O_B}{2}\right) + p (1-f_{B0})^2 -f_{B0}(1-f_{A0}) \right]+ \nonumber\\
    &+ \alpha \left[ (1-2p) \left( \frac{1-f_{B0}+O_B}{2}\right) \left( \frac{1-f_{A0}-O_A}{2}\right) + (1-2p) \left( \frac{1-f_{B0}-O_B}{2}\right) \left( \frac{1-f_{A0}+O_A}{2}\right) \right]+ \nonumber\\
    &+ \alpha \left[p(1-f_{B0})(1-f_{A0})-f_{B0}(1-f_{A0}) \right], \label{eq-dtfbz}
\end{align}
Note that the equations (\ref{eq-dtob}--\ref{eq-dtfbz}) are the same as (\ref{eq-dtoa}--\ref{eq-dtfaz}), except with indices $A$ and $B$ switched, as the system is symmetric.
The detailed derivation is presented in the Appendix.
This set of $4$ differential equations describes how mean opinions and neutral fractions of both groups are evolving in time.
What we are interested in are the stable fixed points of this dynamics.
Due to the algebraic complexity of the equations, finding a general analytical solution is infeasible.
We investigate instead two specific solutions and confirm numerically that no other stable solutions exist.
If all agents belong to a single group, as in \cite{biswas2012}, then for $p<\tfrac{1}{4}$ two symmetric stable states with non-zero opinion $O*$ are stable and for $p>\tfrac{1}{4}$ there exists only one stable solution $O=0$, with all opinions equally present, i.e. $f_0=f_+=f_-=\tfrac{1}{3}$.
By analogy to spin systems in modular environment \cite{suchecki2006, lambiotte2007, suchecki2009}, we consider an antisymmetric ordered state, where each group is internally ordered but with opposite signs.
We expect that in our model a similar situation may occur, and aside from symmetric order states ($O_A>0,~O_B>0$ and $O_A<0$,~$O_B<0$) the antisymmetric states ($O_A>0$,~$O_B<0$ and $O_A<0$,~$O_B>0$) may be stable for low enough levels $p$.
Therefore we reduce the full set of differential equations (\ref{eq-dtoa}--\ref{eq-dtfbz}) to two special cases
\begin{enumerate}
\item Symmetric ordered state $O_B=O_A$ and $f_{A0}=f_{B0}$
\item Antisymmetric ordered state $O_B=-O_A$ and $f_{A0}=f_{B0}$
\end{enumerate}
These two assumptions allow us to drastically reduce the complexity of the system, so that it will be described by only two, simpler differential equations.

\subsection{Symmetric ordered state}
\label{sec-analytics-parallel}
Under the assumption of symmetric state $O_B=O_A$ and $f_{A0}=f_{B0}$, the equation set (\ref{eq-dtoa}--\ref{eq-dtfbz}) reduces to
\begin{align}
    \frac{dO_A}{dt} &=O_A \left[ (1-p) f_{A0} - p \right] \label{eq-p-dtoa}\\
    \frac{df_{A0}}{dt} &= 2 (1-2p) \left(\frac{1-f_{A0}+O_A}{2}\right) \left(\frac{1-f_{A0}-O_A}{2}\right) +\nonumber\\
    &+p(1-f_{A0})^2-f_{A0}(1-f_{A0}), \label{eq-p-dtfaz}
\end{align}
which is independent of the value of the parameter $\alpha$.
These equations are in fact identical to the situation investigated in \cite{biswas2012} when no modular structure is present, and the solutions are the same.
There are four fixed points
\begin{equation}
    \left( O_A^*, f_{A0}^* \right) \in \left\{ (0,\tfrac{1}{3}), (0,1), (O^+,f^+), (-O^+,f^+)\right\}. \label{eq-p-fixed}
\end{equation}
The first two fixed points always exist, while the latter two exist for $p \leq \tfrac{1}{4}$.
The values of $O^+$,$f^+$ given by
\begin{align}
    O^+=\frac{\sqrt{1-4p}}{1-p} \label{eq-p-ostar}\\
    f^+=\frac{p}{1-p} \label{eq-p-fstar}
\end{align}
and in agreement with \cite{biswas2012}.\\
The stability of these fixed points can be analyzed by considering the Jacobian matrix of the original, four-variable equation set (\ref{eq-dtoa}--\ref{eq-dtfbz}) at the fixed points calculated under the symmetric state assumption.
We cannot limit ourselves to considering the stability of the reduced equation set because the enforced equalities $O_B=O_A$ and $f_{A0}=f_{B0}$ could artificially stabilize some fixed points that are, in fact, unstable.
Eigenvalue analysis of the Jacobian matrix shows that the fixed point $(0,\tfrac{1}{3})$ is stable only for $p \geq \tfrac{1}{4}$, the fixed point $(0,1)$ is always unstable and fixed points $(O^+,f^+)$ and $(-O^+,f^+)$ are stable for all values of $p$ where they exist ($p \leq \tfrac{1}{4}$).
Overall, the symmetric ordered state behaves exactly as one group system without a modular structure.

\subsection{Antisymmetric state}
\label{sec-analytics-antiparallel}
Under the assumption of antisymmetric state $O_B=-O_A$ and $f_{A0}=f_{B0}$, the equation set (\ref{eq-dtoa}--\ref{eq-dtfbz}) reduces to a more complicated case than for the symmetric state
\begin{align}
    \frac{dO_A}{dt} &=O_A \left[ (1-p) f_{A0} - \alpha (1-2p) (f_{A0} + 1 ) - p \right] \label{eq-a-dtao}\\
    \frac{df_{A0}}{dt} &= 2 (1-2p) \left(\frac{1-f_{A0}+O_A}{2}\right) \left(\frac{1-f_{A0}-O_A}{2}\right) +\nonumber\\
    &+ \alpha (1-2p) \left[ \left(\frac{1-f_{A0}+O_A}{2}\right) - \left(\frac{1-f_{A0}-O_A}{2}\right) \right]^2 + p(1-f_{A0})^2-f_{A0}(1-f_{A0}) \label{eq-a-dtfaz}
\end{align}
Similar to the symmetric state case, we also find four fixed points
\begin{equation}
    \left( O_A^*, f_{A0}^* \right) \in \left\{ (0,\tfrac{1}{3}), (0,1), (O^-,f^-), (-O^-,f^-)\right\}. \label{eq-a-fixed}
\end{equation}
The first two fixed points always exist, while the second pair $(O^-,f^-)$ and $(-O^-,f^-)$ only exists if
\begin{equation}
    p(1-2\alpha)<\tfrac{1}{4}-\alpha \label{eq-a-condition}
\end{equation}
with the limit $\alpha \rightarrow 0$ resulting in the condition $p \leq \tfrac{1}{4}$ seen in a single group (since it corresponds to the situation of two separate groups).
The values of $O^-$ and $f^-$ are given by formulas
\begin{align}
    O^-=\frac{\sqrt{1-4p-4\alpha(1-2p)}}{\alpha+p-2\alpha p -1} \label{eq-a-ostar}\\
    f^-=-1-\frac{1}{\alpha+p-2\alpha p -1} \label{eq-a-fstar}
\end{align}
that depend on both $p$ and $\alpha$ parameter values, unlike the case of the symmetric state.\\
The stability analysis using the Jacobian matrix of the full equation set (\ref{eq-dtoa}--\ref{eq-dtfbz}) shows that the two fixed points $(0,\tfrac{1}{3})$ and $(0,1)$ behave exactly as for the symmetric case.
The fixed points $(O^-,f^-)$ and $(-O^-,f^-)$ are different.
While they exist for a range of parameters expressed by Eq. \ref{eq-a-condition}, they are stable in a more limited range, only when
\begin{equation}
    p<\frac{7-2\alpha(11-8\alpha)-\sqrt{(3-2\alpha)(3-2\alpha(3-4\alpha)^2)}}{8(1-\alpha)(1-2\alpha)} \label{eq-a-stability}
\end{equation}
if $\alpha \leq \tfrac{1}{6}$.
For $\alpha > \tfrac{1}{6}$, these solutions are unstable for any value of $p$.
This means that the state of two clashing groups internally in consensus loses stability at a lower noise level $p$ than where it ceases to exist.
This results in a discontinuous transition from the antisymmetric state to a symmetric state if noise value $p$ crosses the stability threshold (Eq. \ref{eq-a-stability}).
Similarly, in the case of changing $\alpha$, a discontinuous transition occurs if the inter-group interaction stops fulfilling this stability condition, that can be expressed as
\begin{equation}
    \alpha<\frac{5+2p(6p-11)-\sqrt{4p(p(4p(p-7)+51)-31)+25}}{4(4p(p-2)+3)} \label{eq-a-stabilitya}
\end{equation}
This is the same condition as Eq. \ref{eq-a-stability} but expressed for a different variable.

\section{Numerical results}
\label{sec-results}
\begin{figure}[thb]
\centering
\includegraphics[width=0.75\columnwidth]{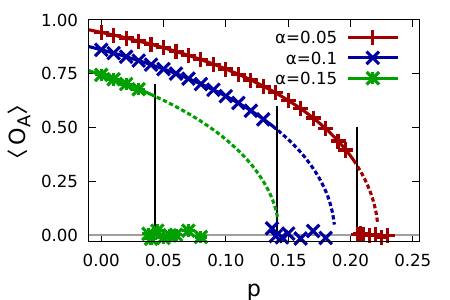}
\caption{The mean opinion $O$ of the system when the initial condition is a fully ordered antisymmetric initial state depends on parameter $p$ according to analytical predictions. The points show the dependence of mean opinion $\langle O_A \rangle$ on the noise level $p$ for a few different intergroup interaction strength $\alpha=0.05$,$\alpha=0.1$ and $\alpha=0.15$ (top to bottom series), while lines display the analytical prediction (Eq. \ref{eq-a-ostar}) when it is stable (solid) and unstable (broken). The vertical line marks the predicted values $p_c(0.05) \approx 0.205$, $p_c(0.1) \approx 0.141$, $p_c(0.015)\approx 0.0434$ (Eq. \ref{eq-a-stability}) where anitsymmetric state stability is lost, which is in a relatively good agreement with the point where the simulated system undergoes the transition from the antisymmetric to a symmetric state (either positive or negative mean opinions of both groups) resulting in average over different realizations close to zero.}
\label{fig-oavsp}
\end{figure}

To verify our analytical approach, we have performed numerical simulations of the BChS model with two groups of $N=2^{14}=16384$ agents each ($N_A=N_B=N$).
The simulations were asynchronous, with each time step consisting of $2N$ updates of single, randomly chosen agent $i$ according to Eq. \ref{eq-oit}.
An agent $j$ to interact with agent $i$ was chosen at random from either different group than agent $i$ (with probability $\alpha$) or from the same group (with probability $1-\alpha$), meaning that in network terms the system corresponds to two annealed complete graphs with fraction $\alpha$ of links rewired to the other group instead of its own in pairs, preserving degrees $k=N-1$ of each node.
Since the symmetric states described in Sec. \ref{sec-analytics-parallel} are essentially the same as for a single, uniform network investigated before \cite{biswas2012}, our simulations focused on the antisymmetric state and its stability.
As the probability for the antisymmetric state to appear from other states due to fluctuations is vanishingly small, our simulations always started from a fully ordered antisymmetric state with $O_A=1$, $O_B=-1$ and $f_{A0}=f_{B0}=0$.
The system was simulated up to 20,000 timesteps to reach an equilibrium state, and this process was repeated 2,000 times to average over different realizations of the stochastic dynamics.
The mean opinion $O_A$ for the equilibrium starting from the antisymmetric ordered initial state has been shown in Fig. \ref{fig-oavsp}

\begin{figure}[thb]
\centering
\includegraphics[width=0.75\columnwidth]{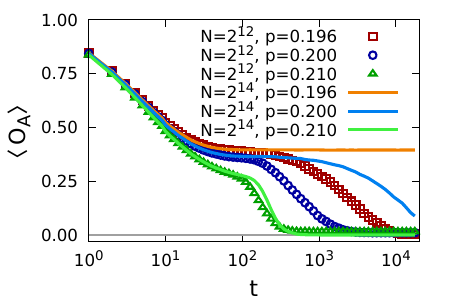}
\caption{The time evolution of measured mean opinion $\langle O_A \rangle$ for systems close to transition point $p_c\approx 0.205$ (Eq. \ref{eq-a-stability}) of different sizes. The plateau indicating metastability of the antisymmetric state is seen for $p<p_c$ and the time it takes to switch to a symmetric state increases with system size, indicating that the discrepancy between analytical and numerical transition point is a finite size effect. In contrast, the time it takes for the transition to a symmetric state to happen above predicted $p_c$ is not increased with the system size.}
\label{fig-oavst}
\end{figure}

The values obtained numerically agree well with the analytical predictions for the stable antisymmetric state expressed by Eq. \ref{eq-a-ostar}.
The mean value $\langle O_A \rangle$ drops to zero at a certain point that is relatively close to the analytically predicted transition point $p_c$ (Eq. \ref{eq-a-stability}) where the antisymmetric state loses its stability.
The drop to zero is a result of the system transition to a symmetric state with both groups either having positive or negative mean opinion with equal probability, which yields a value close to zero after averaging over many realizations.
While our analytical prediction slightly overestimates the threshold noise, which is typical for mean-field approaches, the observed transition point changes with $p$ in the same manner as expected.
The discrepancy is due to the finite size of the system.
Since the antisymmetric state is a metastable state, similar to states corresponding to local energy minima in systems with Hamiltonian, fluctuations can force the transition from the antisymmetric to the symmetric state even below the critical noise level $p_c$, while they never do the opposite.
This causes fluctuations to actively lower required noise $p$ for transition to occur.
We have investigated numerically the behavior of the systems of a different size.
The results, shown in Fig. \ref{fig-oavst} for $\alpha = 0.05$ indicate that transitions to the symmetric state (zero observable $O_A$ due to probabilistic mix of negative and positive symmetric states) for $p=0.196<p_c\approx0.205$ and $p=0.2<p_c$ take significantly more time to occur when the system size increases from $N=2^{12}$ to $N=2^{14}$, so much in fact for $p=0.196$, that we cannot see the signs of transition at all in the $10^4$ simulation steps done.
At the same time, transition for $p=0.21>p_c$, where the system is expected to become symmetric, occurs almost at the same time, regardless of the system size.
The time it takes for the system to complete the transition towards a consensus state becomes shorter closer to the transition point and longer for larger systems.
This shows that the antisymmetric state does present an attractor below the analytically predicted value of $p_c$, with fluctuations switching the system to the symmetric state given enough time for a finite system, while the transition above $p_c$ always occurs quickly.
In conclusion, the lower values of the parameter $p$ at the transition point observed in simulations is the finite size effect driven by fluctuations.

\begin{figure}[thb]
\centering
\includegraphics[width=0.75\columnwidth]{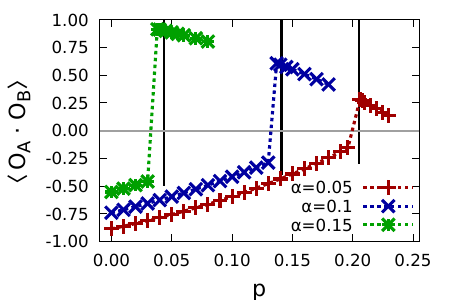}
\caption{When the noise level $p$ is above the predicted critical value $p_c$, the system always switches from the antisymmetric to a symmetric state. The points show the mean product of opinions in both groups $\langle O_A \cdot O_B \rangle$ depending on noise level $p$ for a few different intergroup interaction strengths $\alpha=0.05$, $\alpha=0.1$, and $\alpha=0.15$ (bottom to top series). The vertical lines are the same as in Fig. \ref{fig-oavsp} and mark the predicted values of $p_c(\alpha)$ (Eq. \ref{eq-a-stability}) where the antisymmetric state ($\langle O_A \cdot O_B \rangle<0$) should become unstable and should switch to a symmetric state ($\langle O_A \cdot O_B \rangle >0$).}
\label{fig-oaobvsp}
\end{figure}

To show that the system indeed transits to a symmetric ordered state above $p_c$ and the measured near-zero values of $\langle O_A \rangle$ are not a result of some other phenomena, we also measured the product of both group opinions $\langle O_A \cdot O_B \rangle$, shown in Fig. \ref{fig-oaobvsp}.
As seen in the figure, the product changes from negative values $\langle O_A \cdot O_B \rangle<0$ corresponding to an antisymmetric state to positive ones $\langle O_A \cdot O_B \rangle>0$ corresponding to a symmetric state.\\

\begin{figure}[thb]
\centering
\includegraphics[width=0.75\columnwidth]{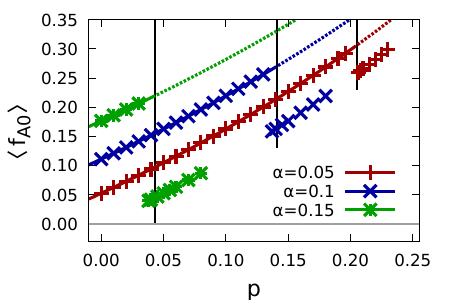}
\caption{The density of neutral opinions in a stable state follows analytical predictions. The points show the density of neutral opinions $\langle f_{A0} \rangle$ depending on the noise level $p$ for three interaction strengths $\alpha=0.05$, $\alpha=0.1$, and $\alpha=0.15$ (bottom to top series), with lines showing the corresponding analytical predictions (Eq. \ref{eq-a-fstar}) when they are stable (solid lines) and unstable (broken lines). The vertical lines (same as Fig. \ref{fig-oavsp} and Fig. \ref{fig-oaobvsp}) show the predicted values of $p_c(\alpha)$ (Eq. \ref{eq-a-stability}) where the antisymmetric state loses stability and undergoes the transition to a symmetric state that is independent of $\alpha$.}
\label{fig-fa0vsa}
\end{figure}

We have also observed the density of neutral opinions $f_{A0}$ in the same situation and show the results in Fig. \ref{fig-fa0vsa}.

The observed neutral opinion density $\langle f_{A0} \rangle$ fits very well to the analytical predictions of Eq. \ref{eq-a-fstar}, up to a point $p_c$ (Eq. \ref{eq-a-stability}) where it drops to a $\alpha$-independent value expected of the symmetric state.
The vertical lines show the predicted transition point $p_c$, indicating again, that it happens in simulations very close to where we expect it to happen.
One interesting thing to note is that as the system approaches the transition, the fraction of undecided agents with neutral opinions grows.
After the transition, when the groups share the same mean opinion, this amount significantly drops to the same level as for independent groups, as the entire system now shares one mean opinion, without the ongoing clash resulting in an increased fraction of agents with neutral opinion.
We have also verified, that mean $f_{A0}=f_{B0}$, so the assumption made during analytical calculations was justified.\\
Overall, the gathered numerical results indicate that the analytical approach presented in Sec. \ref{sec-analytics} correctly describes the behavior of the BChS model, in particular the behavior of the antisymmetric state exclusive to the modular system investigated in this work.

\section{Conclusions}
\label{sec-conclusions}
\begin{figure}[thb]
    \centering
    \includegraphics[width=0.75\columnwidth]{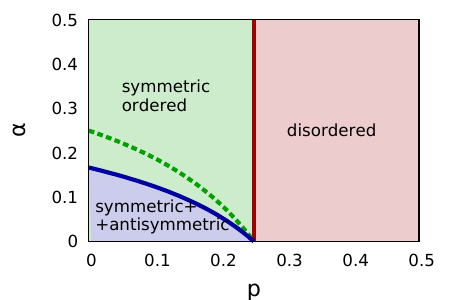}
	\caption{The BChS model in a modular two-group environment can behave in three different ways, depending on the noise level $p$ and the group interconnection frequency $\alpha$, depicted here in colored regions divided by solid lines. For high noise $p>\tfrac{1}{4}$, the only stable state is a disordered state, where each opinion is equally represented, marked in red on the diagram. For $p<\tfrac{1}{4}$ a pair of ordered states is stable, where the mean opinion in both groups is the same and non-zero (Eq. \ref{eq-p-ostar},\ref{eq-p-fstar}). In addition, if $p$ is below critical value $p_c(\alpha)$ expressed by Eq. \ref{eq-a-stability} then an additional pair of antisymmetric states is stable (Eq. \ref{eq-a-ostar},\ref{eq-a-fstar}), where non-zero opinions are opposite in each group. The transition between order and disorder at $p=\tfrac{1}{4}$ (vertical red solid line) is continuous, while the transition between an antisymmetric state allowed for $p<p_c$ to a symmetric state (solid blue curved line) is discontinuous. The transition from the symmetric to the antisymmetric state does not happen naturally. The broken green line represents the limit for the existence of the antisymmetric state (Eq. \ref{eq-a-condition}) and between this line and the solid blue line the antisymmetric state can exist, but it is unstable.}
    \label{fig-phasediagram}
\end{figure}

The BChS model in a modular environment, where agents are divided into two groups with frequent interactions within groups and rare interactions across groups, displays three possible stable states.
Two of these states -- a symmetric ordered state and a disordered state are direct equivalents of states known to exist for a model in a non-modular system, with critical noise level $p_c=\tfrac{1}{4}$ dividing behavior between low-noise ordered state and high-noise disordered state.
For the model in a modular environment, an antisymmetric state can also exist, where each group is internally ordered, but the mean opinions of both groups are opposite.
We have developed an analytical approach that predicts that the state exists and is stable below a certain noise level $p_c$ (Eq. \ref{eq-a-stability}) depending on the strength of the interaction between groups $\alpha$ as long as $\alpha<\tfrac{1}{6}$.
Our theory also predicts the value of the stable mean opinion $O^-$ (Eq. \ref{eq-a-ostar}) of a group, that is, a difference between densities of agents with positive and negative personal opinions $O=f_+-f_-$ as well as the value of stable density of agents with neutral opinion $f^-$ (Eq. \ref{eq-a-fstar}) in this state.
The analytical predictions are confirmed by numerical simulations and can be summarized in a phase diagram for the model, shown in Fig. \ref{fig-phasediagram}.
Although our mean-field approach slightly overestimates the exact transition point $p_c$ for finite systems, it correctly predicts its discontinuous nature and how the transition point changes with other parameter values.

It is worth noting that while the antisymmetric state is stable for $p<p_c$ (Eq. \ref{eq-a-stability}), a symmetric state is also stable and while the transition from antisymmetric to symmetric can be induced by parameter change or noise, the reverse transition is only possible through fluctuations and is extremely unlikely for larger systems.

Stability of simultaneously existing domains of different ordering has been shown to be possible in oscillator synchronization \cite{abrams2004chimera,smirnov2017}, with the related system configurations termed as ``chimera" states.
It is worth noting, however that the nature of such states is different than in our model.
Chimera states can exist in systems with uniform interactions, while in our case, the co-existence of opposite opinions is due to the modular structure of interactions.

Our model may be relevant for two groups following two different ideologies or supporting two different candidates.
The coexistence of the two opposing opinions can persist when the mixing of opinions between communities is limited and when the level of non-conformist behavior signified by $p$ is small.
While such a situation is extremely unlikely to appear due to the evolution of opinions themselves if the topology of interactions is fixed, it may arise as a consequence of changing interaction structure.
It naturally arises when two groups that held prior opposing opinions start interacting.
Another possibility is its emergence through co-evolution between opinions and interaction topology, where people can choose whom they talk with, an effect known as the creation of echo chambers or information bubbles \cite{terren2021, jedrzejewski2020spontaneous}.
The stability of the polarized opinion state in itself means that opinion dynamics present no hindrance to such processes.

\section{Acknowledgement}
KS and JH have received funding from the Polish National Science Center under Alphorn Grant No. 2019/01/Y/ST2/00058, from the European Union under the Horizon Europe grant OMINO (Grant No. 101086321), and from the Polish Ministry of Education and Science with the program International Project Co-Funding.
PS acknowledges hospitality at Warsaw University of Technology and financial support from SERB project (Government of India)  MTR/2020/000356.

\section{Appendix}
This appendix shows the derivation of differential equations (\ref{eq-dtoa}--\ref{eq-dtfaz}) describing the mean-field dynamics of the two-group system.\\
We start with determining the transition rates for each combination of the original and final state $-,0,+$, which represent how many agents of a given opinion change to another opinion during a single update.
\begin{align}
    w_{A++} &= f_{A+} \left[ (1-\alpha)(f_{A0}+(1-p)f_{A+}+p f_{A-})+\alpha (f_{B0}+(1-p)f_{B+}+p f_{B-})\right] \label{aeq-wpp}\\
    w_{A+0} &= f_{A+} \left[ (1-\alpha)((1-p)f_{A-}+p f_{A+})+\alpha ((1-p)f_{B-}+p f_{B+})\right] \label{aeq-wpz}\\
    w_{A+-} &= 0 \label{aeq-wpm}\\
    w_{A0+} &= f_{A0} \left[ (1-\alpha)((1-p)f_{A+}+p f_{A-})+\alpha ((1-p)f_{B+}+p f_{B-})\right] \label{aeq-wzp}\\
    w_{A00} &= f_{A0} \left[ (1-\alpha) f_{A0}+\alpha f_{B0} \right] \label{aeq-wzz}\\
    w_{A0-} &= f_{A0} \left[ (1-\alpha)((1-p)f_{A-}+p f_{A+})+\alpha ((1-p)f_{B-}+p f_{B+})\right] \label{aeq-wzm}\\
    w_{A-+} &= 0 \label{aeq-wmp}\\
    w_{A-0} &= f_{A-} \left[ (1-\alpha)((1-p)f_{A+}+p f_{A-})+\alpha ((1-p)f_{B+}+p f_{B-})\right] \label{aeq-wmz}\\
    w_{A--} &= f_{A-} \left[ (1-\alpha)(f_{A0}+(1-p)f_{A-}+p f_{A+})+\alpha (f_{B0}+(1-p)f_{B-}+p f_{B+}) \label{aeq-wmm}\right]
\end{align}
Note that for the sake of brevity, we only explicitly show transition rates for agents in the group $A$.
Agents in the group $B$ have the same set of transition rates, except for indices $A$ and $B$ being switched.
Each transition rate consists of the probability of the updated agent having a given opinion multiplied by the sum of probabilities of all interaction possibilities that result in the given final opinion.
The product has the following terms:
\begin{enumerate}
\item The probability for the updated agent to have given opinion $o$ is equal to the density of this opinion in the group $f_{Ao}$.
\item The probability of interacting with an agent in the same group $1-\alpha$ or different group $\alpha$
\item The probability for attractive interaction $1-p$ or repulsive interaction $p$
\item The probability for the interacting agent to have opinion $o$ is equal to the density of the given opinion in the same group $f_{Ao}$ or the density of the given opinion in the other group $f_{Bo}$ if it is inter-group interaction
\end{enumerate}
We can write equations for the evolution of the density of each opinion in a group by simply adding up transition rates with factors $+1/N_A$ if the event increases the density of a given opinion, $0$ if it does not impact it, and $-1/N_A$ if the event decreases the density of a given opinion, and at the same time multiplying changes for group $A$ by $N_A$ and group $B$ by $N_B$, since we treat one time step as an update of all agents in the group, hence $N_A$ or $N_B$ individual updates.
The group sizes cancel out, resulting in effective coefficients $+1$, $0$ and $-1$ before the transition rates in the resulting opinion density evolution equations shown below.
\begin{align}
    \frac{df_{A+}}{dt}&=w_{A0+}-w_{A+0}+2 w_{A-+}-2 w_{A+-} \label{aeq-dfap}\\
    \frac{df_{A0}}{dt}&=w_{A0+}+w_{A-0}-w_{A0+}-w_{A0-} \label{aeq-dfaz}\\
    \frac{df_{A-}}{dt}&=w_{A0-}-w_{A-0}+2 w_{A+-}-2 w_{A-+} \label{aeq-dfam}
\end{align}
Like before, we only show equations for group $A$.
There are 3 equivalent equations for group $B$ that only have indices $A$ and $B$ switched.
Putting transition rates (Eq. \ref{aeq-wpp}-\ref{aeq-wmm}) into Eq. (\ref{aeq-dfap}-\ref{aeq-dfam}) gives us proper differential equations for the evolution of the three state densities
\begin{align}
    \frac{df_{A+}}{dt} &= f_{A0} \left[ (1-\alpha) \left( (1-p)f_{A+}+p f_{A-} \right) + \alpha \left( (1-p) f_{B+} + p f_{B-} \right) \right]- \nonumber\\
    &-f_{A+} \left[ (1-\alpha) \left( (1-p) f_{A-} + p f_{A+} \right) + \alpha \left( (1-p) f_{B-} + p f_{B+} \right) \right] \label{aeq-dfap2}\\
    \frac{df_{A0}}{dt} &= f_{A+} \left[ (1-\alpha) \left( (1-p)f_{A-}+p f_{A+} \right) + \alpha \left( (1-p) f_{B-} + p f_{B+}+ \right) \right] \nonumber\\
    &+f_{A-} \left[ (1-\alpha) \left( (1-p)f_{A+}+p f_{A-} \right) + \alpha \left( (1-p) f_{B+} + p f_{B-}+ \right) \right]- \nonumber\\
    &-f_{A0} \left[ (1-\alpha) \left( (1-p)f_{A+}+p f_{A-} \right) + \alpha \left( (1-p) f_{B+} + p f_{B-}+ \right) \right]- \nonumber\\
    &-f_{A0} \left[ (1-\alpha) \left( (1-p)f_{A-}+p f_{A+} \right) + \alpha \left( (1-p) f_{B-} + p f_{B+}+ \right) \right] \label{aeq-dfaz2}\\
    \frac{df_{A-}}{dt} &= f_{A0} \left[ (1-\alpha) \left( (1-p)f_{A-}+p f_{A+} \right) + \alpha \left( (1-p) f_{B-} + p f_{B+} \right) \right]- \nonumber\\
    &-f_{A-} \left[ (1-\alpha) \left( (1-p) f_{A+} + p f_{A-} \right) + \alpha \left( (1-p) f_{B+} + p f_{B-} \right) \right] \label{aeq-dfam2}
\end{align}
with three equivalent equations for group $B$ having exactly same form, but indices $A$ and $B$ switched.
Note that these equations are not independent, but due to the normalization of total densities $f_{A+}+f_{A0}+f_{A-}=1$, $f_{B+}+f_{B0}+f_{B-}=1$ are in fact co-linear.
We reduce the number of variables from the total of $6$ to $4$ to be equal to the number of actual degrees of freedom the system possesses.
Using mean opinions $O_A=f_{A+}-f_{A-}$, $O_B=f_{B+}-f_{B-}$ and densities of neutral opinions $f_{A0}$,$f_{B0}$ as independent variables allows us to write equations for $O_A$ and $O_B$ as
\begin{align}
    \frac{dO_A}{dt}=\frac{d(f_{A+}-f_{A-})}{dt}=\frac{df_{A+}}{dt}-\frac{df_{A-}}{dt} \label{aeq-dtoa}\\
    \frac{dO_B}{dt}=\frac{d(f_{B+}-f_{B-})}{dt}=\frac{df_{B+}}{dt}-\frac{df_{B-}}{dt} \label{aeq-dtob}
\end{align}
which together with equations for $f_{A0}$ (Eq. \ref{aeq-dfaz2}) and $f_{B0}$ (equivalent) express dynamics of the system.
Finally, we can reduce the number of variables and express the dynamics only through $O_A$, $f_{A0}$, $O_B$ and $f_{B0}$.
We put expressions for changes of opinion densities $f_{A+}$, $f_{A-}$, $f_{B+}$, $f_{B-}$ (Eq. \ref{aeq-dfap2}-\ref{aeq-dfam2}) into above equations, substitute all cases of $(f_{A+}-f_{A-})$ as $O_A$, all cases of $(f_{B+}-f_{B-})$ as $O_B$, all cases of $(f_{A+}+f_{A-})$ as $1-f_{A0}$ and all cases of $(f_{B+}+f_{B-})$ as $1-f_{B0}$ and simplify resulting expressions.
Performing these steps gives us the final four equations \ref{eq-dtoa}-\ref{eq-dtfbz} for the dynamics of independent variables, shown and further analyzed in Sec. \ref{sec-analytics}.
\bibliography{main}

\end{document}